\documentclass[12pt,dvips]{article}
\usepackage{epsfig}
\usepackage{color}
\usepackage{latexsym}
\newcommand{\nc}{\newcommand}
\nc{\postscript}[2] 
{\setlength{\epsfxsize}{#2\hsize}\centerline{\epsfbox{#1}}}
\nc{\bg}{B. Grzadkowski}
\nc{\non}{\nonumber}
\nc{\hc}{\hbox {h.c.}} \nc{\re}{\hbox {Re}} 
\nc{\mev}{\hbox {MeV}} \nc{\gev}{\;\hbox {GeV}} \nc{\tev}{\;\hbox {TeV}}
\def\lsim{\mathrel{\raise.3ex\hbox{$<$\kern-.75em\lower1ex\hbox{$\sim$}}}}
\def\gsim{\mathrel{\raise.3ex\hbox{$>$\kern-.75em\lower1ex\hbox{$\sim$}}}}

\nc{\prd}[3]{{\it Phys.\ Rev.}\ {{\bf D{#1}} (#2), #3}}
\nc{\prl}[3]{{\it Phys.\ Rev.\ Lett.}\ {{\bf {#1}} (#2), #3}}
\nc{\plb}[3]{{\it Phys.\ Lett.}\ {{\bf B{#1}} (#2), #3}}
\nc{\npb}[3]{{\it Nucl.\ Phys.}\ {{\bf B{#1}} (#2), #3}}
\nc{\ptp}[3]{{\it Prog.\ Theor.\ Phys.}\ {{\bf {#1}} (#2), #3}}
\nc{\zfp}[3]{{\it Z.\ Phys.}\ {{\bf C{#1}} (#2), #3}}
\nc{\epj}[3]{{\it Eur.\ Phys.\ J.}\ {{\bf C{#1}} (#2), #3}}
\nc{\mpla}[3]{{\it Mod.\ Phys.\ Lett.}\ {{\bf A{#1}} (#2), #3}}
\nc{\rmp}[3]{{\it Rev.\ Mod.\ Phys.}\ {{\bf {#1}} (#2), #3}}
\nc{\ijmpa}[3]{{\it Int.\ J.\ of\ Mod.\ Phys.}\
               {{\bf A{#1}} (#2), #3}}
\nc{\Lsp}{\;\;\;\;\;\;\;\;\;\;}  \nc{\LLLsp}{\lspace \lspace}
\nc{\lsp}{\;\;\;\;\;\;}
\nc{\spac}{\;\;\;}
\nc{\noi}{\noindent}
\nc{\beq}{\begin{equation}}   \nc{\eeq}{\end{equation}}
\nc{\bea}{\begin{eqnarray}}   \nc{\eea}{\end{eqnarray}}
\nc{\baa}{\begin{array}}      \nc{\eaa}{\end{array}}
\nc{\bit}{\begin{itemize}}    \nc{\eit}{\end{itemize}}
\nc{\ben}{\begin{enumerate}}  \nc{\een}{\end{enumerate}}
\nc{\bce}{\begin{center}}     \nc{\ece}{\end{center}}

\def\phio{\phi_0}
\def\mho{m_{h_0}}
\def\mphio{m_{\phi_0}}
\def\fbi{~\mbox{fb$^{-1}$}}
\def\ie{{\it i.e.}}
\def\anti{\overline}
\def\mz{m_Z}
\def\mw{m_W}
\def\wp{W^+}
\def\wm{W^-}
\def\gam{\gamma}
\def\h{h}
\def\mh{m_{h}}
\def\mphi{m_\phi}
\def\what{\widehat}
\def\lwh{\widehat\Lambda_W}

\def\lphi{\Lambda_\phi}

\def\mphi{m_\phi}

\def\hbar{\overline h}
\def\lam{\lambda}
\def\mpl{M_{Pl}}

\def\ifmath#1{\relax\ifmmode #1\else $#1$\fi}
\def\half{\ifmath{{\textstyle{1 \over 2}}}}

\def\quarter{\ifmath{{\textstyle{1 \over 4}}}}

\def\call{{\cal L}}

\def\eps{\epsilon}

\def\vev#1{\langle #1\rangle} 
\nc{\tb}{\stackrel{{\scriptscriptstyle (-)}}{t}}
\nc{\bb}{\stackrel{{\scriptscriptstyle (-)}}{b}}
\nc{\fb}{\stackrel{{\scriptscriptstyle (-)}}{f}}
\nc{\pp}{\gamma \gamma}
\nc{\pptt}{\pp \to \ttbar}
\nc{\barh}{\overline{h}}
   \def\thebibliography#1{\centerline{REFERENCES}
     \list{[\arabic{enumi}]}{\settowidth\labelwidth{[#1]}\leftmargin
     \labelwidth\advance\leftmargin\labelsep\usecounter{enumi}}
     \def\newblock{\hskip .11em plus .33em minus -.07em}\sloppy
     \clubpenalty4000\widowpenalty4000\sfcode`\.=1000\relax}
   
\begin{document}
\vspace*{-2.5cm}
\begin{flushright}
$\vcenter{
\hbox{IFT-02-18}
\hbox{UCD-02-09} 
\hbox{DFF-389-06-02}
%\hbox{hep-ph/0206197}
\hbox{June, 2002}
}$
\end{flushright}
\renewcommand{\thefootnote} {\alph{footnote})}
\bce 
{\large\bf HIGGS-BOSON INTERACTIONS WITHIN THE RANDALL-SUNDRUM MODEL}\footnote{
Talk presented by J.~F.~Gunion at the 
``Planck 02'',   Fifth European Meeting  From the Planck Scale to the Electroweak Scale, 
``Supersymmetry and Brane Worlds'',   Kazimierz, Poland, May 25 - 29, 2002}
\vspace*{.8cm}

{\sc Daniele DOMINICI$^{\:1),\:}$}\footnote{E-mail address:
\tt dominici@fi.infn.it}
{\sc Bohdan GRZADKOWSKI$^{\:2),\:}$}\footnote{E-mail address:
\tt bohdan.grzadkowski@fuw.edu.pl}\\
{\sc John F. GUNION$^{\:3),\:}$}\footnote{E-mail address:
\tt jfgucd@higgs.ucdavis.edu} {\sc Manuel TOHARIA$^{\:3),\:}$}\footnote{E-mail address:
\tt toharia@physics.ucdavis.edu}
\vspace*{.8cm}

{\sl
 $^1$ Dipartimento di Fisica, Florence University and INFN, \\
 Via Sansone 1, 50019 Sesto. F. (FI), ITALY\\
\vskip 0.1cm
$^2$ Institute of Theoretical Physics, Warsaw 
University,\\
 Ho\.za 69, PL-00-681 Warsaw, POLAND\\
\vskip 0.1cm
$^3$ Davis Institute for High Energy Physics, 
University of California Davis, \\ Davis, CA 95616-8677, USA \\
}
\vskip .8cm
Dedicated to Stefan Pokorski on his 60th birthday.\\
{\it To appear in Acta Physica Polonica B}
\ece
\vskip .8cm
\centerline{ABSTRACT} 
\vspace*{0.2cm} \baselineskip=14pt plus 0.1pt minus 0.1pt 
The scalar sector of the Randall-Sundrum
model is discussed.  The effective potential for the Standard Model
Higgs-boson ($h$) interacting with Kaluza-Klein excitations of the
graviton ($h_\mu^{\nu\, n}$) and the radion ($\phi$) has been derived 
and it has been shown that
{\it only} the Standard Model vacuum solution of $\partial V/\partial
h =0$ is allowed.  The theoretical and experimental consequences of the
curvature-scalar mixing $\xi\, R\, \what H^\dagger \what H$ 
introduced on the visible brane are considered
and simple sum rules that relate the couplings of the mass eigenstates $h$ and $\phi$
to pairs of vector bosons and fermions are derived.
The sum rule for the $ZZh$ and $ZZ\phi$ couplings 
in combination with LEP/LEP2 data implies that 
not both the $h$ and $\phi$ can be light. We present explicit
results for the still allowed region in the $(\mh,\mphi)$ plane 
that remains after
imposing the LEP upper limits for non-standard scalar couplings to a
$ZZ$ pair. 
The phenomenological consequences of the mixing are investigated and,
in particular, it is shown that 
the Higgs-boson decay $h \to \phi\phi$ would provide an
experimental signature for non-zero $\xi$ and can have 
a very substantial impact on the Higgs-boson searches, 
having $BR(h \to \phi\phi)$
as large as $30 \div 40\, \%$. 

\vskip .3cm



%--------------------------------------------------------------------
\renewcommand{\thefootnote}{\arabic{footnote}}
%--------------------------------------------------------------------
\pagestyle{plain} \setcounter{footnote}{0}
\baselineskip=20.6pt plus 0.2pt minus 0.1pt

%%%%%%%%%%%%%%%%%%%%%%%%%%%%%%%%%%%%%%%%%%%%%%%%%%%%%%%%%%%%%%%%%%%%%%%%

\section{Introduction}

Although the Standard Model (SM) of electroweak interactions describes
successfully almost all existing experimental data the model
suffers from many theoretical drawbacks. One of these is the hierarchy
problem: namely, the SM can not consistently accommodate the weak
energy scale ${\cal O}(1\tev)$ and a much higher scale such as the
Planck mass scale ${\cal O}(10^{19}\gev)$.  Therefore, it is commonly
believed that the SM is only an effective theory emerging as the
low-energy limit of some more fundamental high-scale theory that
presumably could contain gravitational interactions.  In the last few
years there have been many models proposed that involve extra
dimensions. These models have received tremendous attention since they
could provide a solution to the hierarchy problem. One of the most
attractive attempts has been formulated by Randall and
Sundrum~\cite{rs}, who postulated a 5D universe with two 4D surfaces
(``3-branes''). All the SM particles and forces with the exception of
gravity are assumed to be confined to one of those 3-branes called the
visible brane.  Gravity lives on the visible brane, on the second
brane (the ``hidden brane'') and in the bulk.  All mass scales in the
5D theory are of the order of the Planck mass.  By placing the SM
fields on the visible brane, all the  mass terms 
(of the order of the Planck mass) are
rescaled by an exponential suppression factor (the ``warp factor'')
$\Omega_0\equiv e^{-m_0 b_0/2}$, which reduces them down to the weak
scale ${\cal O}(1 \tev)$ on the visible brane without any severe fine
tuning. To achieve the necessary suppression, one needs $m_0 b_0 /2
\sim 35$. This is a great improvement compared to the original problem
of accommodating both the weak and the Planck scale within a single
theory.

In order to obtain a consistent solution to the Einstein
equations corresponding to a low-energy effective theory that is flat,
the branes must have
equal but opposite cosmological constants and these must
be precisely related to the bulk cosmological constant.

The model is defined by the 5D action:
\bea
S&=&-\int d^4x\, dy \sqrt{-\what g}\left({R\over2\what\kappa^2}+\Lambda\right)\\
&&+\int d^4x\,\sqrt{-g_{hid}}({\cal L}_{hid}-V_{hid})
+\int d^4x\,\sqrt{-g_{vis}}({\cal L}_{vis}-V_{vis}),\non
\label{action}
\eea
where $\what g^{\what\mu\what\nu}$ ($\what\mu,\what\nu=0,1,2,3,4$)
is the bulk metric and 
$g_{hid}^{\mu\nu}(x)\equiv \what g^{\mu \nu}(x,y=0)$ and
$g_{vis}^{\mu\nu}(x)\equiv \what g^{\mu \nu}(x,y=1/2)$ ($\mu,\nu=0,1,2,3$)
are the induced metrics on the branes.
One finds that if the bulk and brane
cosmological constants are related by
$\Lambda/m_0=-V_{hid}=V_{vis}=-6m_0/\hat{\kappa}^2$ and if 
periodic boundary conditions identifying $(x,y)$ with $(x,-y)$ are imposed, 
then the 5D Einstein equations lead to the following metric:
\beq
ds^2=e^{-2\sigma(y)}\eta_{\mu\nu}dx^\mu dx^\nu-b_0^2dy^2, 
\label{metricz}
\eeq
where
$\sigma(y)=m_0b_0\left[y(2\theta(y)-1)-2(y-1/2)\theta(y-1/2)\right]$;
$b_0$ is a constant parameter that is not determined by the 
action, Eq.~(\ref{action}).
Gravitational fluctuations around the above background metric
will be defined through the replacement:
\beq
\eta_{\mu\nu} \to \eta_{\mu\nu}+\epsilon h_{\mu\nu}(x,y)\lsp b_0\to b_0+b(x)\,.
\label{metric}
\eeq
Below we will be expanding in powers of $\eps h_{\mu\nu} =
\sqrt{2}\hat{\kappa}h_{\mu\nu}$ and $b(x)/b_0$.

The paper is organized as follows. First, 
in Sec.~\ref{secpotential} we derive 
the effective potential
for the SM Higgs-boson sector interacting with  
Kaluza-Klein excitations of the graviton  ($h_\mu^{\nu\, n}$)
and the radion ($\phi$). 
In Sec.~\ref{secmixing}, we introduce  
the curvature-scalar mixing $\xi\, R\, \what H^\dagger \what H$  and discuss its
consequences for couplings and interactions. 
In Sec.~\ref{secpheno}, we discuss some  phenomenological aspects
of the scalar sector, focusing on the particularly important
possibility of $h\to\phi\phi$ decays. 
We summarize our results in Sec.~\ref{secfinal}.

Our work extends in several ways the already extensive
literature \cite{Bae:2000pk,Davoudiasl:1999jd,Cheung:2000rw,Davoudiasl:2000wi,Han:2001xs,Park:2000xp,Chaichian:2001rq,Hewett:2002nk}
on the phenomenology of the Randall-Sundrum model. We focus
in particular on the case where the radion is substantially
lighter than the Higgs boson and the important impacts of
Higgs-radion mixing in this case.

%%%%%%%%%%%%%%%%%%%%%%%%%%%%%%%%%%%%%%%%%%%%%
\section{The effective potential}
\label{secpotential}

The canonically normalized
massless radion field $\phio(x)$ is defined by: 
\beq \phio (x) \equiv
\left({6 \over \what\kappa^2 m_0}\right)^{1/2}\Omega_b(x)=\left({6 \over \what\kappa^2 m_0}\right)^{1/2}e^{-m_0 b(x)/2}\,.  
\label{radiondef}
\eeq 

Keeping in mind that $h_{\mu\nu}(x,y)$ depends both on $x$ and $y$, we  
use the KK expansion in the extra dimension 
\beq
h_{\mu\nu}(x,y)=\sum_n h_{\mu\nu}^n(x){\chi^n(y)\over \sqrt b_0}\,.
\label{kk}
\eeq

The total 4D effective potential (up to the  terms of the order of ${\cal O}[(\eps h_{\mu\nu})^3]$) 
has been determined in Ref.\cite{DGGT}. Restricting ourselves
to the trace part of 
$h_{\mu\nu}^n \sim \quarter 
\eta_{\mu\nu}\hbar^n$ the result is the following:
\bea
&&V_{eff}=V_{eff}^{brane}+V_{eff}^{KK}=\non\\
&&\left(1+{1\over\widehat\Lambda_W}
\sum_n\hbar^n+{1\over 4\widehat\Lambda_W^2}\sum_n\sum_m\hbar^n\hbar^m+\cdots\right)\left[
\left(1+{\phio\over\Lambda_\phi}
\right)^4 V(h_0)+\half \mphio^2\phio^2\right]\non\\
&&-{3\over 16}\sum_n m_n^2 (\hbar^n)^2+\cdots\,,
\label{vefftot}
\eea
where $m_n$ is the KK-graviton mass, $\lwh \equiv 2 \sqrt{b_0}/[\eps \chi^n(1/2)]\simeq
\sqrt{2} \mpl \Omega_0$ and we have expanded  around the vacuum expectation
values for the radion, $\phio \to \langle\phi_0\rangle + \phio \equiv \lphi + \phio $.  
In order to stabilize the size of the extra dimension we have introduced the radion mass
$\mphio$ without specifying its origin.
Restricting ourself to the perturbative regime we will look for the 
minimum of $V_{eff}$ that satisfies
$\sum_n\hbar^n/\widehat\Lambda_W\ll 1$ and $b(x)/b_0\ll 1$,
the latter being equivalent to $\phio(x)/\lphi \ll 1$:
\bea
\left({1\over\lwh}+\frac{1} {2 \lwh^2}\sum_n\hbar^n\right)
\left[ \left(1+{\phio\over\Lambda_\phi} 
\right)^4V +\frac12 m_\phi^2\phio^2\right] -{3\over 8} m_n^2\hbar^n&=&0
\non\\ \label{min1}\\
\left(1+{1\over \widehat\Lambda_W}\sum_n\hbar^n+
{1\over 4\widehat\Lambda_W^2}\sum_n\sum_m\hbar^n\hbar^m\right)
4\left(1+{\phio\over\Lambda_\phi}\right)^3 {V\over \Lambda_\phi} +\mphio^2\phio&=&0
\non\\
\label{min2}
\\
\left(1+{1\over \widehat\Lambda_W}\sum_n\hbar^n+
{1\over 4\widehat\Lambda_W^2}\sum_n\sum_m\hbar^n\hbar^m
\right) \left(1+{\phio\over\Lambda_\phi}\right)^4
{\partial V(h_0)\over \partial h_0}&=&0\,.
\non\\
\label{min3}
\eea
There is only one solution of Eq.~(\ref{min3})
consistent with $\phio/\lphi\ll 1$ and $\hbar^n/\lwh\ll 1$: 
namely, ${\partial V(h_0)\over\partial h_0}=0$. 
For consistency of the RS model we must also require that $V(\vev{h_0})=0$.
If $V(\vev{h_0})\neq 0$, then the visible brane tension would
be shifted away from the very finely tuned RS solution to the Einstein
equations. With these two ingredients, Eq.~(\ref{min2})
implies that $\vev{\phio}=0$ at the minimum, implying that we
have chosen the correct expansion point for $\phio$, and
Eq.~(\ref{min1}) then implies that $\vev{\hbar^n}=0$, \ie\
we have expanded about the correct point in the $\hbar^n$ fields.
However, it is only if $\mphio^2>0$ that $\vev{\phio}=0$ is required
by the minimization conditions.
If $\mphio=0$, then Eq.~(\ref{min1}) still requires $\vev{\hbar^n}=0$
but all equations are satisfied for any $\vev{\phio}$.

Finally, we note that since $\partial V/ \partial h_0 =0$ at the minimum
(even after including interactions with the radion and KK gravitons)
there are no terms in the potential that are linear in
the Higgs field $h_0$ (so in particular no $h_0-\phi_0$ mass mixing emerge).  
We will return to this observation in the next
section of the paper.

%%%%%%%%%%%%%%%%%%%%%%%%%%%%%%%%%%%%%%%%%%%%
\section{The curvature-scalar mixing}
\label{secmixing}

Having determined the vacuum structure of the model,
we are in a position to discuss the
possibility of mixing between gravity and the electroweak sector.
The simplest example of
the mixing is described by the following action~\cite{johum}: 
\beq
S_\xi=\xi \int d^4 x \sqrt{g_{\rm vis}}R(g_{\rm vis})\what H^\dagger \what H\,,
\eeq
where $R(g_{\rm vis})$ is the Ricci scalar for the metric induced on the visible brane
$g^{\mu\nu}_{\rm vis}=\Omega_b^2(x)(\eta^{\mu\nu}+\eps h^{\mu\nu})$.
Using $H_0=\Omega_0 \what H$ one obtains~\cite{csaki_mix}
\beq
\xi\sqrt{g_{\rm vis}}R(g_{\rm vis})\what H^\dagger \what H=6\xi\Omega(x)\left(-\Box\Omega(x)+
\eps h_{\mu\nu}\partial^\mu
\partial^\nu \Omega + \cdots\right)H_0^\dagger H_0\,.
\label{ksiphi}
\eeq
To isolate the kinetic energy terms we use the expansion 
\beq
H_0={1\over \sqrt 2}(v_0+h_0)\,,\quad \Omega(x)=1+{\phi_0\over \lphi}\,.
\label{expansion}
\eeq
The $h_{\mu\nu}$ term of Eq.~(\ref{ksiphi}) 
does not contribute to the kinetic energy
since a partial integration would lead to 
$h_{\mu\nu}\partial^\mu\partial^\nu \Omega=
 -\partial^\mu h_{\mu\nu}\partial^\nu\Omega=0$ by virtue
 of the gauge choice, $\partial^\mu h_{\mu\nu}=0$. 
We thus find the following kinetic energy terms:
\beq
\call=-\half\left\{1+6\gamma^2 \xi \right\}\phi_0\Box\phi_0
-\half\phi_0 \mphio^2\phi_0-\half h_0 (\Box+\mho^2)h_0-6\gamma \xi \phi_0\Box h_0\,,
\label{keform}
\eeq
where $\gamma\equiv v_0/\lphi$ and $\mho^2=2\lam v^2$ and
$\mphio^2$ are the Higgs and radion masses before the mixing.
The above differs from Ref.~\cite{wells_mix} by the extra $\phi_0\Box\phi_0$ 
piece proportional to $\xi$.

We define the mixing angle $\theta$ by
\beq
\tan 2\theta\equiv 12 \gam \xi Z {\mho^2\over \mphio^2-\mho^2(Z^2-36\xi^2\gam^2)}\,,
\label{theta}
\eeq
where
\beq
Z^2\equiv 1+6\xi\gam^2(1-6\xi)\,.
\label{z2}
\eeq
In terms of these quantities, the states that diagonalize the kinetic energy
and have canonical normalization are $h$ and $\phi$ with: 
\bea
h_0&=&\left (\cos\theta -{6\xi\gam\over Z}\sin\theta\right)h
+\left(\sin\theta+{6\xi\gam\over Z}\cos\theta\right)\phi\equiv d h+c\phi
\label{hform}\\
\phi_0&=&-\cos\theta {\phi\over Z}+\sin\theta {h\over Z}\equiv a\phi+bh\,. \label{phiform}
\eea
(Our sign convention for $\phi_0$ is opposite to that chosen for $r$
in Ref.~{\cite{csaki_mix}.)
To maintain positive definite kinetic energy terms for the $h$ and $\phi$,
we must have $Z^2>0$.  (Note that this implies that 
$\beta\equiv 1+6\xi \gam^2>0$ is implicitly required.)
The corresponding mass-squared eigenvalues are
\beq
m_\pm^2={1\over 2 Z^2}\left(\mphio^2+\beta \mho^2\pm\left\{
[\mphio^2+\beta \mho^2]^2-4Z^2\mphio^2\mho^2\right\}^{1/2}\right)\,.
\label{emasses}
\eeq
It follows from the above formula that
$m_\pm$ cannot be too close to being degenerate
in mass, depending on the precise values of $\xi$ and $\gam$, see Ref.\cite{DGGT}.

We now turn to the important interactions of the $h$, $\phi$ and 
$h_{\mu\nu}^n$. We begin with the $ZZ$ couplings of the $h$ and $\phi$.
The $h_0$ has standard $ZZ$ coupling while the $\phi_0$ has $ZZ$
coupling deriving from the interaction $-{\phi_0\over \lphi}T_\mu^\mu$
using the covariant derivative  portions of $T_\mu^\mu(h_0)$.  The result 
for the $\eta_{\mu\nu}$ portion of the $ZZ$ couplings is:
\beq
g_{ZZh}=\frac{g\,\mz}{c_W}\left(d+\gamma b\right), \lsp 
g_{ZZ\phi}=\frac{g\,\mz}{c_W}\left(c+\gamma a\right)\,,
\eeq
where $a$, $b$, $c$ and $d$ are defined 
through Eqs.~(\ref{hform},\ref{phiform}) and $g$,  $c_W$ denote the $SU(2)$ gauge coupling 
and cosine of the Weinberg angle, respectively . 
The $WW$ couplings are obtained by replacing $g\mz/c_W$ by $g\mw$.
Notice also an absence of $Zh\phi$ tree level couplings.
Next, we consider the fermionic couplings of the $h$ and $\phi$.
The $h_0$ has standard fermionic couplings and the fermionic
couplings of the $\phi_0$ derive from $-{\phi_0\over \lphi}T_\mu^\mu$
using the Yukawa interaction contributions to $T_\mu^\mu$.
One obtains results in close analogy to the $VV$ couplings just
considered:
\beq 
g_{f\bar{f}h}=-\frac{g\,m_f}{2\,\mw}(d+\gamma b) \lsp 
g_{f\bar{f}\phi}=-\frac{g\,m_f}{2\,\mw}(c+\gamma a)
\label{yuk}
\eeq

For small values of $\gam$, the $g_{ZZh}$ and $g_{ZZ\phi}$ 
have the expansions: 
\bea
g_{ZZh}&=&\left(\frac{g\,\mz}{c_W}\right)\left[1+{\cal O}(\gamma^2)\right], \\
g_{ZZ\phi}&=&\left(\frac{g\,\mz}{c_W}\right) 
\left[-\gamma\left(1+\frac{6\xi m_\phi^2}{\mh^2-m_\phi^2}\right) +
{\cal O}(\gamma^3)\right]\non  
\eea
Entirely analogous results apply for the fermionic couplings. 

The following simple and exact sum rules (independently noted
in \cite{Han:2001xs}) follow from the definitions
of $a,b,c,d$:
\beq
{g_{ZZh}^2+g_{ZZ\phi}^2\over \left(\frac{g\,\mz}{c_W}\right)^2}=
{g_{f\anti f h}^2+g_{f\anti f \phi}^2\over \left(\frac{gm_f}{2\mw}\right)^2}=
\left[1+{\gamma^2(1-6\xi)^2\over Z^2}\right]\equiv R^2\,.
\label{summ_Z}
\eeq 
Note that $R^2>1$ is a result of the
non-orthogonality of the relations Eq.~(\ref{hform}) and Eq.~(\ref{phiform}). 
Of course, $R^2=1$ in the conformal limit, $\xi=1/6$.
It is important to note that $Z\to 0$ 
would lead to divergent $ZZ$ and $f\anti f$ couplings for the $\phi$.
As noted earlier, this was to be anticipated since $Z\to 0$ corresponds to 
vanishing of the radion kinetic term before going to canonical
normalization. 
After the rescaling that guarantees the canonical normalization, 
if $Z\to 0$ the radion coupling constants blow up: 
$g_{ZZ\phi} \propto (c+\gamma a) \simeq 1/(6 \xi \gamma Z) + {\cal O}(Z)$. 
To have $Z^2>0$, $\xi$ must lie in the region:
\beq
\frac{1}{12}\left(1-\sqrt{1+\frac{4}{\gamma^2}}\right)
\leq \xi \leq
\frac{1}{12}\left(1+\sqrt{1+\frac{4}{\gamma^2}}\right)\,.
\label{xilim}
\eeq
As an example, for $\lphi=5\tev$, $Z^2>0$ corresponds to the range
$-3.31\leq\xi\leq 3.47$.  Of course, if we choose $\xi$ sufficiently
close to the limits, $Z^2\to 0$ implies that the couplings, as
characterized by $R^2$ will become very large.  Thus, we 
should impose bounds on $\xi$ that keep $R^2$ moderate in size.
For example, for $\lphi=5\tev$,  
$R^2$ in Eq.~(\ref{summ_Z}) takes the values 2.48 and 1.96  at $\xi=-2.5$
and $\xi=2.5$, respectively.  Therefore we will impose an overall
restriction of $R\leq 5$. In practice, this bound seldom plays a role, being
almost always superseded by the constraint limiting $\xi$ according to the degree of $m_h-m_\phi$
degeneracy.

The final crucial ingredient for the phenomenology that we shall consider
is the tri-linear interactions among the $h$ and $\phi$ and $h_{\mu\nu}^n$
fields.  In particular, these are crucial for 
the decays of these three types of particles. The tri-linear interactions
derive from four basic sources.

\begin{enumerate}
\item
First, we have the cubic interactions coming from 
\beq
\call\ni -V(H_0)= - \lambda (H_0^\dagger H_0 - \frac 1 2 v_0^2)^2=
-\lambda  (v_0^2 h_0^2 + v_0 h_0^3 + \frac 1 4 h_0^4)\,,
\label{vhexpansion}
\eeq
after substituting $H_0=\frac {1} {\sqrt{2}} (v_0 + h_0)$.
Since $\lambda$ is related to the bare Higgs-boson mass
in a usual way: $\mho^2=2\lambda v_0^2$  
the $h_0^3$ interaction can be expressed as
\beq
\call \ni -{\mho^2\over 2 v_0}h_0^3\,.
\eeq
\item
Second, there is the interaction of the radion 
$\phi_0$ with the stress-energy momentum tensor trace:
\beq
\call\ni -{\phi_0\over\lphi}T_\mu^\mu(h_0)=-{\phi_0\over\lphi}\left(-\partial^\rho h_0\partial_\rho h_0 + 4\lam v_0^2h_0^2\right)\,.
\eeq
\item
Thirdly, we have the interaction of the KK-gravitons with the 
contribution to the stress-energy
momentum tensor coming from the $h_0$ field:
\beq
\call\ni -{\eps\over 2}
 h_{\mu\nu}T^{\mu\nu}\ni -{1\over\lwh}\sum_n h_{\mu\nu}^n 
\partial^\mu h_0\partial^\nu h_0
\,,
\label{kkhiggs}
\eeq
where we have kept only the derivative contributions and
we have dropped (using the gauge $h_\mu^{\mu\,n}=0$) 
the $\eta^{\mu\nu}$ parts of $T^{\mu\nu}$.
\item
Finally, we have
the $\xi$-dependent tri-linear components of Eq.~(\ref{ksiphi}):
\bea
&&6\xi\Omega(x)\left(-\Box\Omega(x)+
\eps h_{\mu\nu}\partial^\mu
\partial^\nu \Omega(x)\right)H_0^\dagger H_0\ni\Bigl[- 3 \frac  \xi \lphi h_0^2 \Box \phi_0 
\non\\
&&- 6 \xi \frac {v_0} {\lphi^2}
h_0 \phi_0 \Box \phi_0 - 12 \xi \frac {v_0}{\lwh \lphi}
\sum_n h_{\mu\nu}^n\partial^\mu\phi_0\partial^\nu h_0
\non\\
&&-6 \xi \frac {v_0^2}{\lwh \lphi^2}
\sum_n h_{\mu\nu}^n\partial^\mu\phi_0\partial^\nu \phi_0\Bigr]
\label{ksitri}
\eea
where we have  employed $\partial^\mu h_{\mu\nu}^n=0$, used
the traceless gauge condition $h_\mu^{\mu\, n}=0$, and also
used the symmetry of $h_{\mu\nu}$. 
\end{enumerate}

As seen from the above list, without 
the curvature-Higgs mixing the lagrangian does not contain 
any interactions linear in the Higgs field, therefore vertices
like  $\phi^2 h$ and $h^n\phi h$ (that follows from Eq.~(\ref{ksitri}))
are a clear indication for the curvature-Higgs mixing.
As we shall see, the  $\phi^2 h$ coupling could also be of considerable phenomenological
importance leading to $h\to \phi\phi$ decays. The $h^n\phi h$ coupling would 
be relevant for  $h\to \phi h^n$, however for the parameters range considered here
that decay would be relatively rare.

%%%%%%%%%%%%%%%%%%%%%%%%%%%%%%%%%%%%%%%%%%%%%
\section{Phenomenology}
\label{secpheno}

We begin by discussing the
restrictions on the $h,\phi$ sector imposed by LEP Higgs-boson searches.
Since no scalar boson (s) was observed in the process $e^+e^-\to Z s$,
LEP/LEP2 provides an upper limit for the coupling of a $ZZ$ pair to the scalar
as a function of the scalar mass. Here we will
employ the limits  from \cite{Abbiendi:1998rd,teixeiradias,:2001xw}.

The first question that arises is whether {\it both} the $\phi$ and the $h$
could be light without either having been detected at LEP and LEP2.
The sum rule of Eq.~(\ref{summ_Z}) implies that this is impossible
since the couplings of
the $h$ and $\phi$ to $ZZ$ cannot both be suppressed.  
For any given value of $\mh$ and $\mphi$, the range of $\xi$ is limited
by: (a) the constraint limiting $\xi$ according to the degree of $m_h-m_\phi$
degeneracy;
(b) the constraint that $Z^2>0$, Eq.~(\ref{z2}); and (c)
the requirements that $g_{ZZh}^2$ and $g_{ZZ\phi}^2$ both lie
below any relevant LEP/LEP2 limit.  The regions in the $(\xi,\mphi)$ plane
consistent with the first two constraints as well as with $R<5$ are shown in
Figs.~\ref{allowed_mh112} and \ref{allowed_mh120} for $\mh=112\gev$
and $\mh=120\gev$, respectively, 
assuming a value of $\lphi=5\tev$.  
For the most part, it is the degeneracy constraint (a)
that defines the theoretically acceptable regions shown.
The regions within the theoretically acceptable
regions that are excluded by the LEP/LEP2
limits are shown by the yellow  shaded regions,
while the allowed regions are in blue.
For $\mh=112\gev$, the LEP/LEP2 limits
exclude a large portion of the theoretically consistent parameter space.
For $\mh=110\gev$ (not plotted), the sum rule of Eq.~(\ref{summ_Z})
results in all of the
theoretically allowed parameter space being excluded by LEP/LEP2 constraints.
For $\mh=120\gev$, the LEP/LEP2 limits do not apply to the $\h$
and it is only for $\mphi\lsim 115\gev$ and significant $g_{ZZ\phi}^2$
(requiring large $|\xi|$) that some points are ruled out
by the LEP/LEP2 constraints. As a result, the allowed region is dramatically
larger than for $\mh=112\gev$.
The precise regions shown are somewhat sensitive to the $\lphi$ choice,
but the overall picture is always similar to that presented here
for $\lphi=5\tev$. 
\begin{figure}[h!]
\begin{center}
\vspace*{-.5cm}
\epsfig{figure=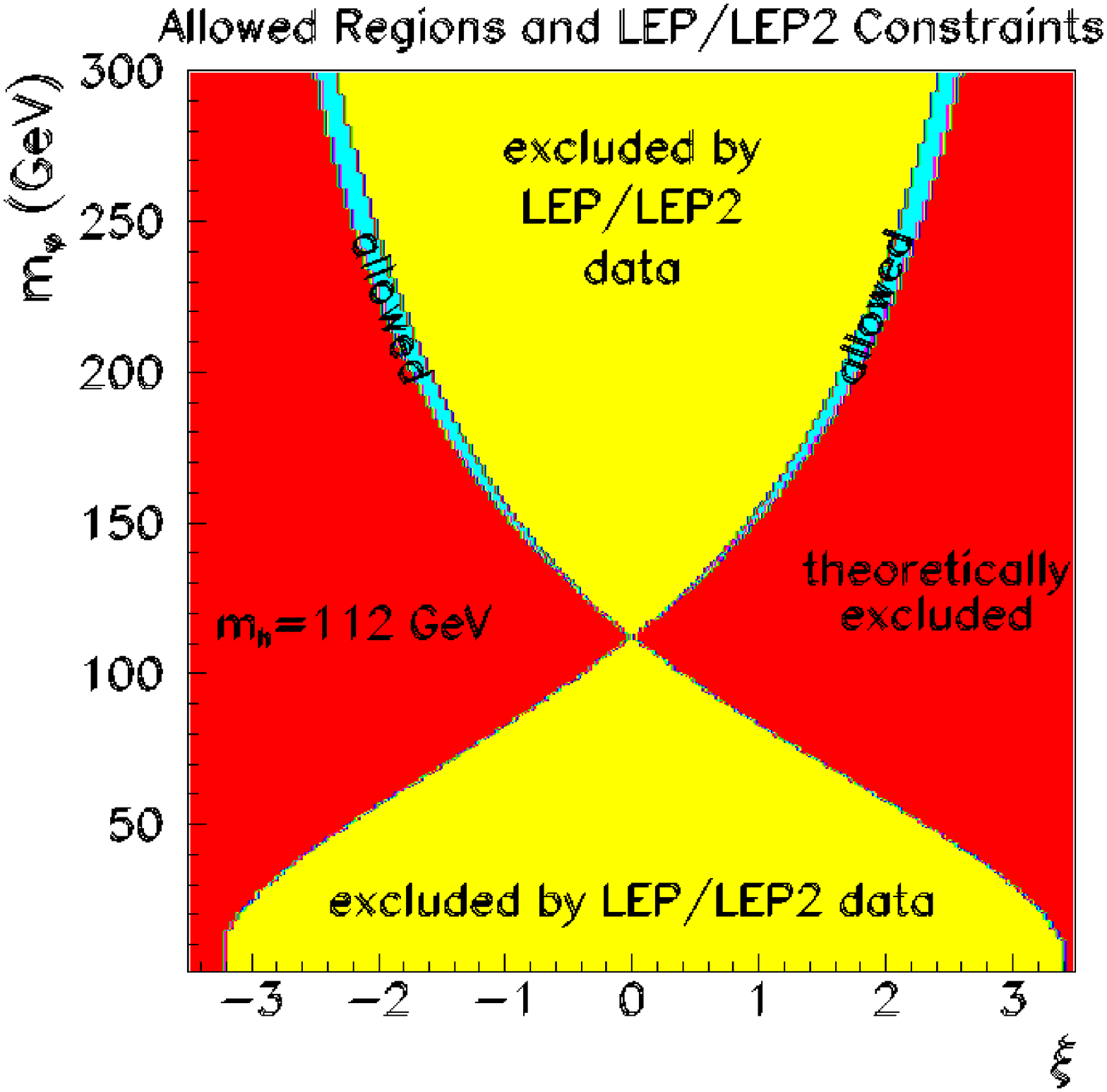,height=8.5cm,width=14.cm}
\end{center}
\vspace*{-.9cm}
\caption{Allowed regions (see text)
in $(\xi,\mphi)$ parameter space for $\lphi=5\tev$
and $\mh=112\gev$. The dark red portion of parameter
space is theoretically disallowed. The light yellow  portion
is eliminated by LEP/LEP2 constraints
on the $ZZs$ coupling-squared $g_{ZZs}^2$ or on
$g_{ZZs}^2BR(s\to b\anti b)$, with $s=\h$ or $s=\phi$. 
}
\label{allowed_mh112}
\vspace*{-.2cm}
\begin{center}
\epsfig{figure=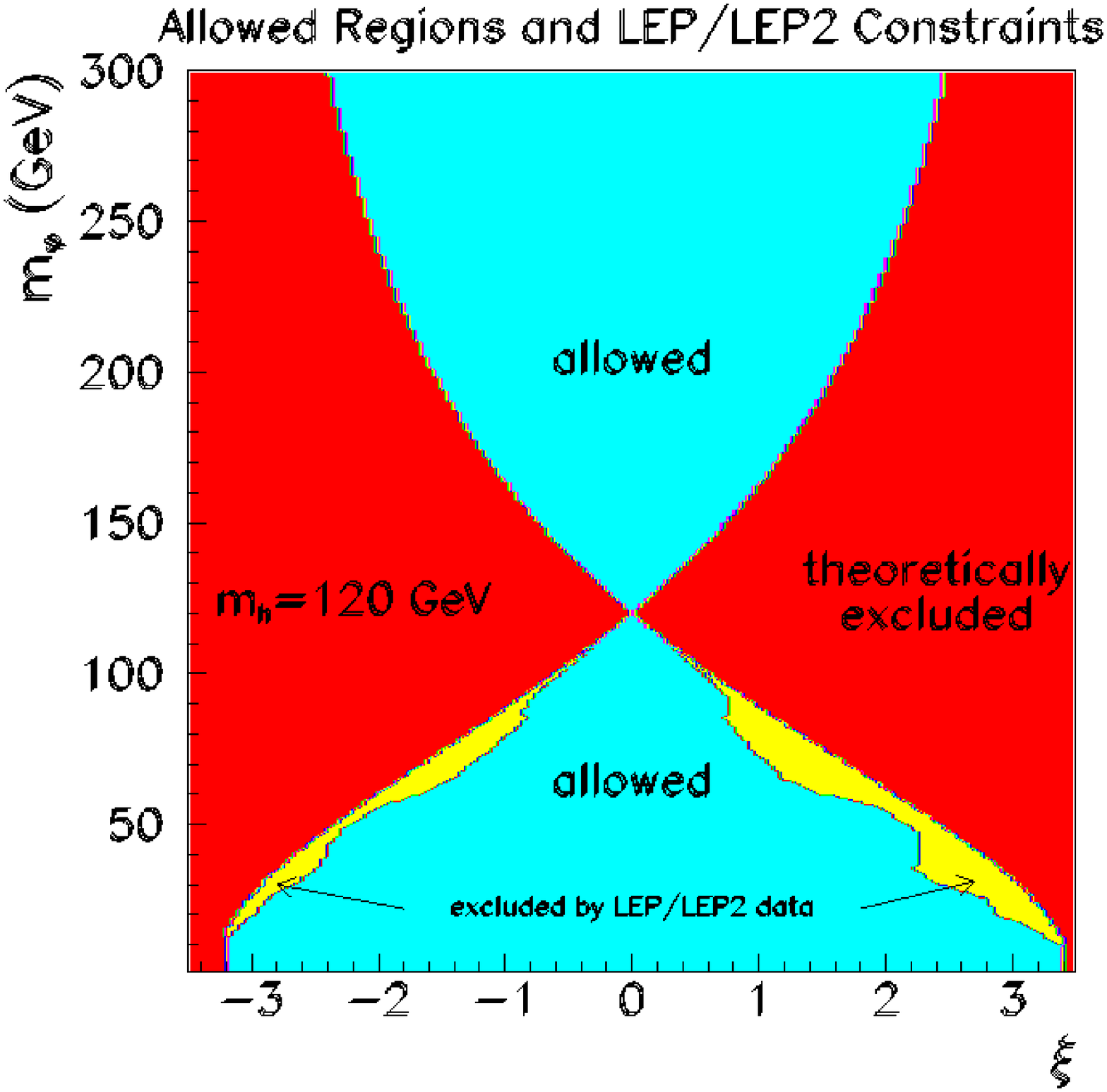,height=8.5cm,width=14cm}
\end{center}
\vspace*{-.9cm}
\caption{As in Fig.~\ref{allowed_mh112} but for
$\mh=120\gev$. 
}
\label{allowed_mh120}
\end{figure}

In order to illustrate LHC Higgs-boson discovery potential in the 
presence of the 
curvature-mixing we plot in Fig.~\ref{prodh_mh120} the
ratio of the rates for $gg\to h \to \gamma\gamma$, $WW\to h \to \tau^+\tau^-$
and $gg\to t\anti t h \to t\anti t b\anti b$ (the latter
two ratios being equal) to the corresponding rates for
the SM Higgs boson. All the curves are plotted for the parameter range that
is consistent with the theoretical and experimental constraints mentioned above.
For this figure, we take $\mh=120\gev$
and $\lphi=5\tev$ and show results for $\mphi=20$, $55$ and $200\gev$.
As will be discussed later,
in the case of $\mphi=55\gev$, the $h\to \phi\phi$ decay is substantial for large $|\xi|$.  
The resulting suppression of the standard LHC modes at the largest allowed
$|\xi|$ values is most evident in the $\wp\wm\to h\to \tau^+\tau^-$
curves. Another important impact of mixing is through 
communication of the anomalous $gg$ coupling
of the $\phi_0$ to the $h$ mass eigenstate. The result is
that prospects for $h$ discovery in the 
$gg\to h \to \gamma\gamma$ mode could be either substantially poorer 
or substantially better than for a SM
Higgs boson of the same mass, depending on $\xi$ and $\mphi$.

\begin{figure}[h]
\begin{center}
\epsfig{figure=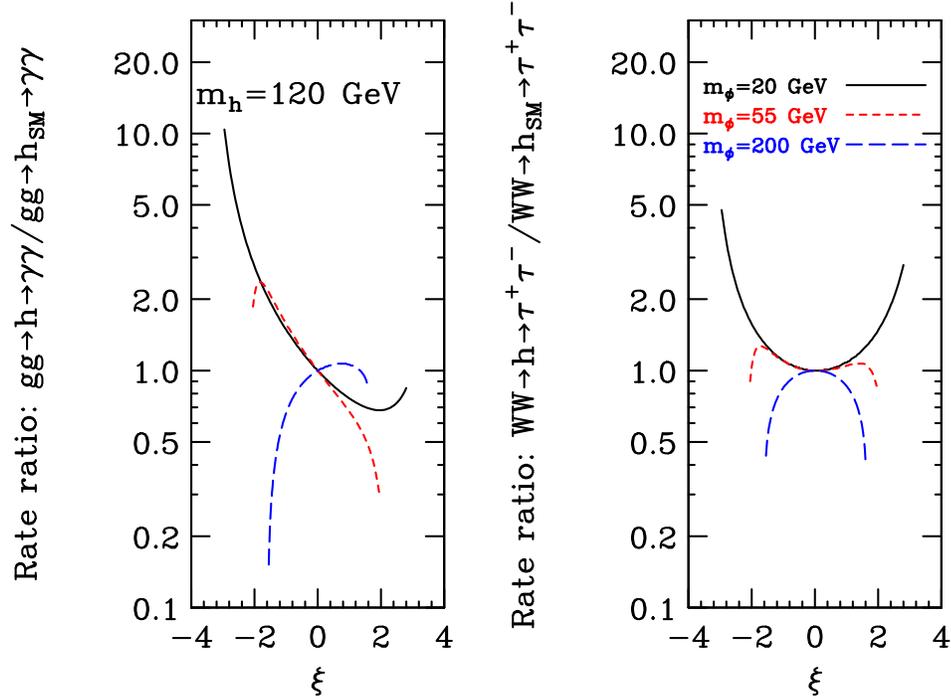,height=10.cm,angle=90,width=12.6cm}
\end{center}
%\vspace*{-.2in}
\caption{The ratio of the rates for $gg\to h \to \gamma\gamma$
and $WW\to h \to \tau^+\tau^-$ (the latter is the same as
that for $gg\to t\anti t h\to t\anti t b\anti b$)
to the corresponding rates for the SM Higgs boson.
Results are shown
for $m_h=120\gev$  and $\lphi=5\tev$ as functions of $\xi$
for $\mphi=20$, $55$ and $200\gev$.
}
\label{prodh_mh120}
\end{figure}

At the LC, the potential for $h$ discovery is primarily determined
by $g_{ZZh}^2$. As we have shown in Ref.~\cite{DGGT}, this 
coupling-squared  (relative to the SM value)
is often $>1$ (and can be as large as $\sim 5$), but can also
fall to values as low as $\sim 0.4$, implying significant
suppression relative to SM expectations.
However the latter suppression is still well within the reach of 
the $e^+e^-\to Z h$ recoil mass discovery technique at a LC with
$\sqrt s=500\gev$ and $L=500\fbi$. 

A particularly important feature of Figs.~\ref{allowed_mh112} and \ref{allowed_mh120} is that
once $\mh$ is large enough (typically $\mh\gsim 115 \gev$
is sufficient) it will generally be possible
to find $\xi$ values for which a range of moderately small,
and possibly even very small, $\mphi$ values
cannot be excluded by LEP/LEP2 constraints. In particular, $\mphi<\mh/2$ (so
that $h\to\phi\phi$ decays are possible) is typically not excluded
for a substantial range of $\xi$. (The reverse is also true;
allowed parameter regions exist for which $\phi\to hh$ decays
are possible once $\mphi\gsim 100\gev$. However,
for this paper we have chosen to focus on cases 
in which the $\phi$ is not very heavy.)
With this in mind, we now turn to a discussion of branching ratios,
focusing on the $h\to \phi\phi$ final mode:
\beq \Gamma(h\to \phi\phi )={g_{\phi\phi h}^2\over
32\pi\mh\lphi^2}(1-4r_\phi)^{1/2}\,, 
\label{htophiphiwdth}
\eeq
where
$\lam(1,r_1,r_2)\equiv 1+r_1^2+r_2^2-2r_1-2r_2-2r_1r_2$,
$r_\phi=\mphi^2/\mh^2$, $r_n=m_n^2/\mh^2$ and\footnote{Note that both 
diagonal physical masses and the bare Higgs-mass parameter $\mho$ appear below.}
\bea
&&g_{\phi\phi h}\equiv
2m_\phi^2\left[6a\xi(\gam(ad+bc)+cd)+bc^2\right]\\
&&+m_h^2 \;c\left[12ab\gam\xi+2ad+bc(6\xi-1)\right]
-4c(2ad+bc)\mho^2 -3\gam^{-1}c^2 d\mho^2\, .\non
\eea

\begin{figure}[h!]
\begin{center}
\epsfig{figure=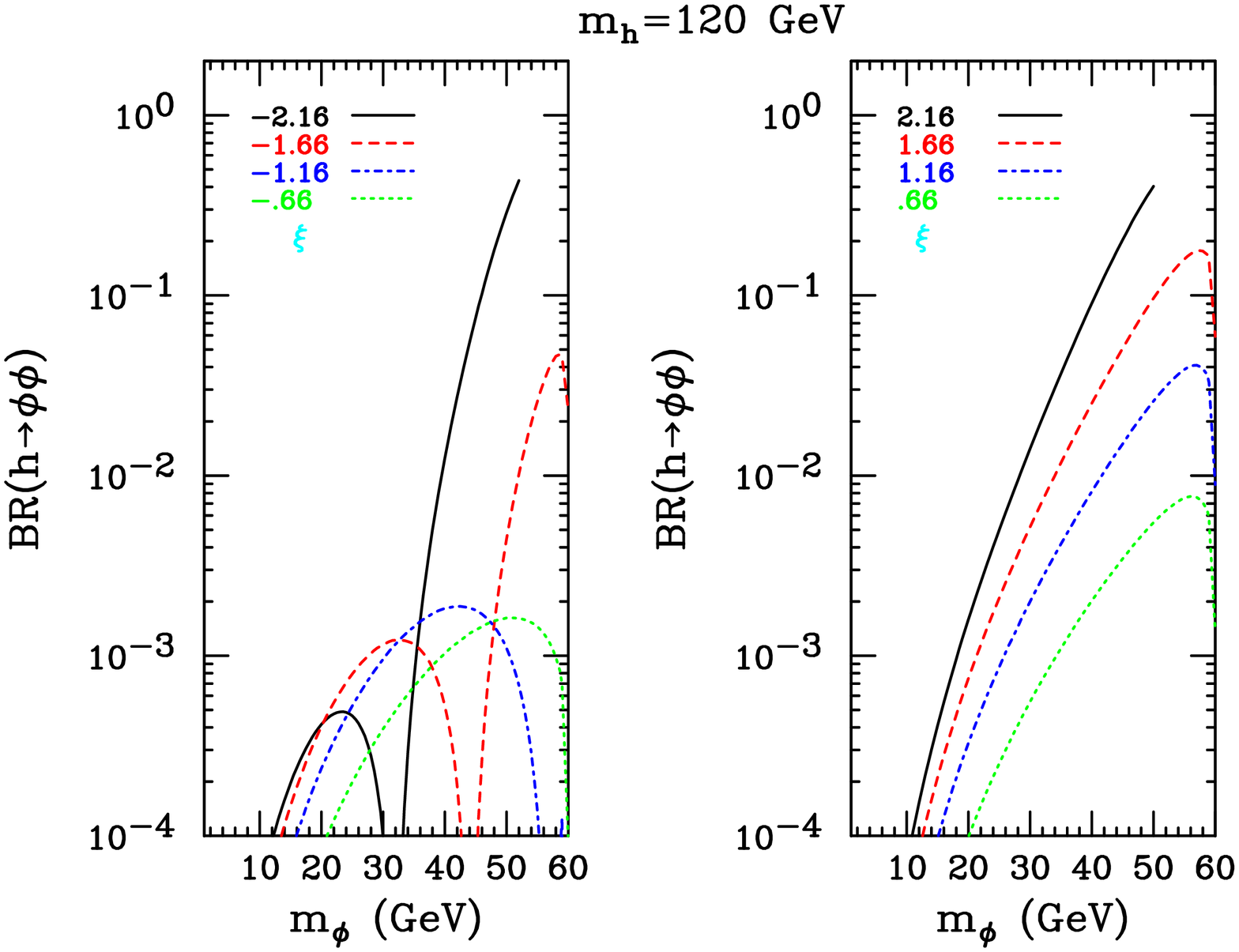,height=10.cm,angle=90,width=10.cm}
\vspace*{-.25cm}
\epsfig{figure=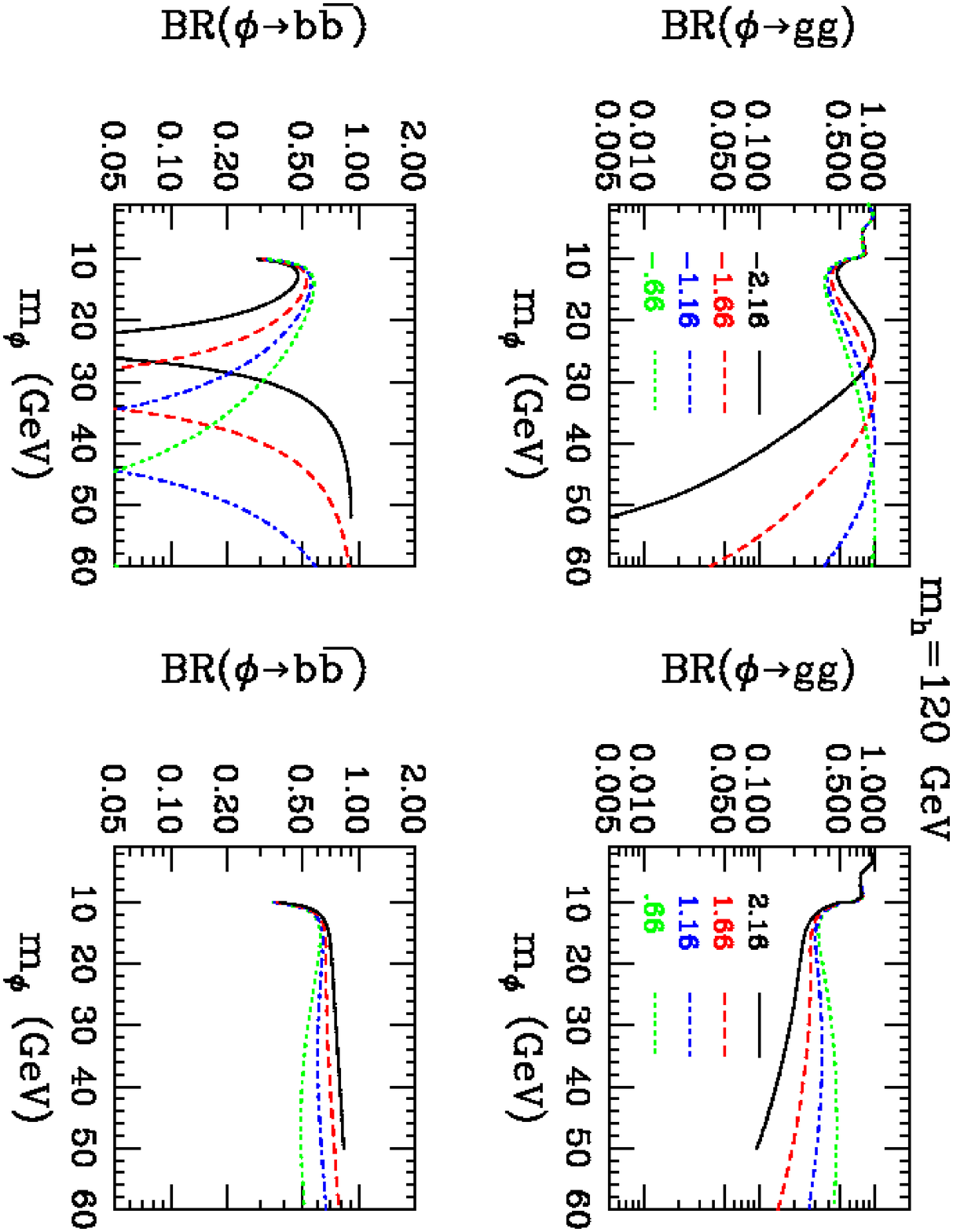,height=10.cm,angle=90,width=10.cm}
\end{center}
\vspace*{-.25cm}
\caption{
The branching ratios for $h\to \phi\phi$, $\phi\to gg$
and $\phi\to b\anti b$ for $m_h=120\gev$  
and $\lphi=5\tev$ as a function of $\mphi$ for $\xi=-2.16$, $-1.66$,
$-1.16$ and $-0.66$ (left-hand graphs) and for $\xi=0.66$, $1.16$,
$1.66$, and $2.16$ (right-hand graphs).
}
\label{br_mh120}
\end{figure}

The branching ratios for $h\to \phi\phi$ in the case of $\mh=120\gev$
and $\lphi=5\tev$ are shown in Fig.~\ref{br_mh120} for various $\xi$
choices within the allowed region. The plots show that $h\to\phi\phi$
decays can be quite important at the largest $|\xi|$ values
when $\mphi$ is close to $\mh/2$.
Detection of the $h\to \phi\phi$
decay mode could easily provide the most striking evidence
for the presence of $\xi\neq 0$ mixing. 
In order to understand how to search for the $h\to\phi\phi$ decay
mode, it is useful to know how the $\phi$ decays.  
In Fig.~\ref{br_mh120} we give detailed results
for $BR(\phi\to gg)$ and $BR(\phi\to b\anti b)$
for the same $\mphi$ and $\xi$ values
for which $BR(h\to \phi\phi)$ is plotted.
(The $c\anti c$ and $\tau^+\tau^-$ channels supply the remainder.)
For $\xi>0$, $BR(\phi\to b\anti b)$ is always substantial and
might make detection of the $\h\to \phi\phi\to 4b$
and $\h\to \phi\phi\to 2g 2b$ final
states possible.
The $\phi\to\gam\gam$ decay mode always has a 
very tiny branching ratio and the related detection channels
would not be useful.

One will probably first search for the $h$ in the modes that
have been shown to be viable for the SM Higgs boson.
We have given in Fig.~\ref{prodh_mh120} the rates for important
LHC discovery modes relative to the corresponding SM values
in the case of $\mphi=55\gev$. Results for other $\mphi<\mh/2$ values
are similar in nature. We observe that
the $WW\to h\to \tau^+\tau^-$ and $gg\to t\anti t h\to t\anti t b\anti b$
detection modes are generally sufficiently mildly suppressed
that detection of the $h$ in these modes should be possible
(assuming full $L=300\fbi$ luminosity per detector).
The $gg\to h\to \gam\gam$
detection mode could either be enhanced or significantly suppressed
relative to the SM expectation.
Once the $h$ has been detected in one of the SM modes, 
a dedicated search for
the $h\to \phi\phi\to b\anti b b\anti b$
and $h\to \phi\phi\to b\anti b gg$ decay modes will be important.
At the LHC, backgrounds for these modes will be substantial
and a thorough Monte Carlo assessment will be needed.

%%%%%%%%%%%%%%%%%%%%%%%%%%%%%%%%%%%%%%%%%%%%%
\section{Summary and Conclusions}
\label{secfinal}

We have discussed the scalar sector of the Randall-Sundrum model. The
effective potential (defined as a set of interaction terms that
contain no derivatives) for the Standard Model Higgs-boson sector
interacting with Kaluza-Klein excitations of the graviton
($h_\mu^{\mu\, n}$) field and the radion ($\phi$) field has been
derived. Without specifying its origin, a stabilizing mass-term for
the radion has been introduced. After including this term, 
we have shown that only the
Standard Model vacuum determined by $\partial V/\partial h =0$ is
allowed. 
An important requisite property for the correct vacuum solution is that
the effective potential does not
contain any terms linear in the Higgs field.

Having confirmed that the usually assumed vacuum properties are 
correct, we pursue in more detail the phenomenology of the RS scalar
sector, focusing in particular on results found in the presence of 
a curvature-scalar mixing $\xi\, R\, \what H^\dagger \what H$ contribution to the Lagrangian.
Simple sum rules that relate Higgs-boson and radion couplings to pairs
of vector bosons and fermions have been derived. 
Of particular interest is the fact that non-zero $\xi$ induces interactions
linear in the Higgs field: $\phi^2 h$ and $h^n h \phi$.

We derive the regions of parameter
space that are excluded by direct LEP/LEP2 limits 
on scalar particles with $ZZ$ coupling as function of scalar mass.
Of particular note is the fact that
the sum rule for $ZZh$ and $ZZ\phi$ squared-couplings noted above 
implies that it is impossible for both the $h$ and $\phi$ to be light. 
However, even very light $\phi$ ($\mphi<10\gev$) remains a possibility
if $\mh\gsim 115\gev$
and the (dominant) $\phi\to gg$ decays result in final states to which
existing searches for on-shell $Z\to Z^*\phi$ decays would not 
have been very sensitive. 

One particularly interesting complication for $\xi\neq 0$ is the presence of 
the non-standard decay channels, $h \to \phi\phi$ and $h \to
h^n \phi$. These could easily be present since in the context of the RS model
there is a possibility (perhaps even a slight preference)
for the $\phi$ to be substantially lighter than the $h$.  
In particular, $\mphi<\mh/2$ is a distinct possibility.
We study in detail the phenomenology when $\mphi\leq 60\gev$
for $\mh=120\gev$,  so that the $h\to\phi\phi$ mode is present. Even
for a relatively conservative choice of the new-physics scale, $\lphi=5\tev$,
this mode will be present at an observable level, and, 
at the largest $|\xi|$ values and for $\mphi$ not far below $\mh/2$, can even 
substantially dilute the rates for the usual $h$ search channels. 
In any case, detection of $h\to \phi\phi$ is very important as it
would provide a crucial experimental signature for
non-zero $\xi$. For the less conservative choice of $\lphi=1\tev$
and for a light $h$, {\it e.g.} $\mh=120\gev$, $BR(h \to \phi\phi)$
could easily be as large as $50\, \%$ for most of the 
theoretically allowed
values of $\mphi$ (which are near $\mh/2$) when $|\xi|$ is near the largest
value allowed by theoretical and existing experimental constraints.

\vspace*{0.6cm}
% AAAAAAAAAAAAAAAAAAAAAAAAAAAAAAAAAAAAAAAAA
\centerline{ACKNOWLEDGMENTS}

\vspace*{0.3cm}

The authors are grateful to the organizers of 
the Fifth European Meeting  From the Planck Scale to the Electroweak Scale, 
``Supersymmetry and Brane Worlds''  for creating a very  warm and 
inspiring atmosphere during the meeting.
B.G. thanks Z.~Lalak, K.~Meissner and J.~Pawelczyk for useful discussions. 
J.F.G would like to thank J. Wells for useful discussions.
B.G. is supported in part by the State Committee for Scientific
Research under grant 5~P03B~121~20 (Poland). J.F.G. is supported
by the U.S. Department of Energy and by the Davis Institute for High
Energy Physics.

\end{document}